\pdfoutput=1

\documentclass[%
 aip,
 jmp,%
 amsmath,amssymb,
 reprint,%
]{revtex4-1}

\usepackage{graphicx}
\usepackage{dcolumn}
\usepackage{ amssymb }
\usepackage{hyperref}

\begin{document}

\title{Disk and elliptical galaxies within renormalization group improved gravity}

\author{Davi C. Rodrigues$^a$}\email{davirodrigues.ufes@gmail.com}  \author{Paulo L. C. de Oliveira$^a$} \author{J\'ulio C. Fabris$^a$}   \author{Ilya L. Shapiro$^b$}
 \affiliation{$^a$ Departamento de F\'{\i}sica, Universidade Federal do Esp\'{\i}rito Santo,  29075-910, Vit\'oria, ES, Brazil \\ $^b$Departamento de F\'{\i}sica, Universidade Federal de Juiz de Fora,  36036-330, Juiz de Fora, MG, Brazil}

\begin{abstract}
The paper is about possible effects of infrared quantum contributions to General Relativity on  disk and elliptical galaxies. The Renormalization Group corrected General Relativity (RGGR model) is used to parametrize these quantum effects. The new RGGR results presented here concern the elliptical galaxy  NGC 4374 and  the dwarf  disk galaxy DDO 47. Using the effective approach to Quantum Field Theory in curved background, one can argue that the proper RG energy scale, in the weak field limit, should be related to the Newtonian potential. In the  context of galaxies, this led to a remarkably small variation of the gravitational coupling G, while also capable of generating galaxy rotation and dispersion curves of similar quality to the the best dark matter profiles (i.e., the profiles that have  a core).\footnote{This paper is based on a talk given by D.C. Rodrigues at the I CosmoSul meeting (Rio de Janeiro, RJ - Brazil. August, 01-05, 2011).}

 \end{abstract}

\maketitle

\renewcommand{\vec}[1]{{\bf #1}}
\renewcommand{\Re}{\,\mbox{Re}\,}
\renewcommand{\Im}{\,\mbox{Im}\,}

\def\<{\left \langle}
\def\>{\right \rangle}
\def\[{\left\lbrack}
\def\]{\right\rbrack}
\def\({\left(}
\def\){\right)}
\newcommand{\be}{\begin{equation}}
\newcommand{\ee}{\end{equation}}
\newcommand{\ea}{\end{eqnarray}}
\newcommand{\ba}{\begin{eqnarray}}
\newcommand{\prt}{{\partial}}
\newcommand{\diag}{\mbox{diag}}
\newcommand{\tr}{\mbox{tr}}
\newcommand{\grad}{\ensuremath{\vec{\nabla}}}
\newcommand{\bs}{\begin{sideways}}
\newcommand{\es}{\end{sideways}}
\newcommand{\chir}{{\chi^2_{\mbox{\tiny{red}}}}} 
\newcommand{\Newt}{N}
\newcommand{\Iso}{{\mbox{\tiny Iso}}}
\newcommand{\mond}{{\mbox{\tiny MOND}}}
\newcommand{\stvg}{{\mbox{\tiny STVG}}}
\newcommand{\IsoInf}{{\mbox{\tiny Iso} \infty}}
\newcommand{\dg}{\dagger} 
\newcommand{\ML}{$\Upsilon_*^V/ (\frac{M_\odot}{L_{\odot,\mbox{\tiny V}}})$}
\newcommand{\amond}{$\frac{a_0}{1.35\times10^{-8} \mbox{\tiny cm/s}^2}$}
\newcommand{\mnras}{Mon. Not. R. Astron. Soc.}
\newcommand{\aap}{Astronomy $\&$ Astrophysics}
\newcommand{\apjs}{ApJS}
\newcommand{\apjl}{Astrophys. J. Letters}
\newcommand{\aj}{Astron. J.}
\newcommand{\pasa}{PASA}
\newcommand{\RGGR}{{\mbox{\tiny RGGR}}}
\newcommand{\MOND}{{\mbox{\tiny MOND}}}
\newcommand{\Ser}{{\mbox{\tiny S}}}
\newcommand{\ext}{{\mbox{\tiny ext}}}

\def\beq{\begin{eqnarray}}
\def\eeq{\end{eqnarray}}
\def\ln{\,\mbox{ln}\,}
\def\Det{\,\mbox{Det}\,}
\def\det{\,\mbox{det}\,}
\def\tr{\,\mbox{tr}\,}
\def\diag{\,\mbox{diag}\,}
\def\Tr{\,\mbox{Tr}\,}
\def\sTr{\,\mbox{sTr}\,}
\def\Res{\,\mbox{Res}\,}

\def\lap{\Delta}
\def\sla{\!\!\!\slash}
\def\al{\alpha}
\def\bet{\beta}
\def\ch{\chi}
\def\ga{\gamma}
\def\de{\delta}
\def\vp{\varepsilon}
\def\ep{\epsilon}
\def\ze{\zeta}
\def\io{\iota}
\def\ka{\kappa}
\def\la{\lambda}
\def\na{\nabla}
\def\pa{\partial}
\def\ro{\varrho}
\def\rh{\rho}
\def\si{\sigma}
\def\om{\omega}
\def\ph{\varphi}
\def\ta{\tau}
\def\th{\theta}
\def\te{\vartheta}
\def\up{\upsilon}
\def\Ga{\Gamma}
\def\De{\Delta}
\def\La{\Lambda}
\def\Si{\Sigma}
\def\Om{\Omega}
\def\Te{\Theta}
\def\Th{\Theta}
\def\Up{\Upsilon}

\section{Introduction} \label{intro}

In Refs. \cite{Rodrigues:2009vf, Rodrigues:2012qm, Rodrigues:2011cq, Farina:2011me, Fabris:2012wg} we presented new results on the application of renormalization group (RG) corrections to General Relativity in the astrophysical domain, in particular on a possible relation between RG large scale effects and dark matter-like effects in galaxies. The resulting phenomenological model was named RGGR. These developments were directly based on the RG application to gravity of Ref. \cite{Shapiro:2004ch}, and are consistent with the phenomenological consequences of diverse approaches to the subject, including the related to the asymptotic safety scenario of quantum gravity, see in particular Refs.  \cite{Reuter:2004nx,Reuter:2004nv,Reuter:2007de}.

Currently, in the context of quantum field theory in curved space time, it is impossible to construct a formal proof that the coupling parameter $G$ is a running parameter in the infrared. However, this possibility can not be ruled out.  The possibility of General Relativity being modified in the far infrared due to the renormalization group (RG) has been considered in different contexts, for instance,  \cite{Goldman:1992qs, Bertolami:1993mh, Dalvit:1994gf, Bertolami:1995rt}. The previous attempts to apply this picture to galaxies have considered for simplicity point-like galaxies (e.g., \cite{Shapiro:2004ch, Reuter:2004nx}). We extended previous considerations by identifying a proper renormalization group energy scale $\mu$  and by evaluating the consequences considering the observational data of disk \cite{Rodrigues:2009vf} and elliptical \cite{Rodrigues:2012qm} galaxies. We proposed in Ref. \cite{Rodrigues:2009vf} the existence of a relation between $\mu$ and the local value of the Newtonian  potential (this relation was reinforced  afterwards \cite{Domazet:2010bk}). With this choice, the renormalization group-based approach (RGGR) was capable to mimic dark matter effects with great precision. Also, it is remarkable that this picture induces a very small variation on the gravitational coupling parameter $G$, namely a variation of about $10^{-7}$ of its value across a galaxy (depending on the matter distribution). We call our model RGGR, in reference to Renormalization Group corrected General Relativity.

Here we present a brief review of the RGGR achievements in galaxies and present new results on the galaxies NGC 4374 and DDO 47. The first is a giant elliptical galaxy that was first analyzed in Ref. \cite{Rodrigues:2012qm} in the context of RGGR and MOND, the second is a well known dwarf disk galaxy whose results within RGGR are for the first time here presented, and it is part of a larger work yet to be published \cite{FabrisFuture}. It should be pointed that disk and elliptical galaxies behave as stationary systems that are stable due to different physical reasons, disk galaxies are essentially axisymmetric bodies supported by rotation,  while elliptical galaxies are about spherically symmetric and mainly supported by velocity dispersions. Hence models for galaxy kinematics may in principle succeed in, say, disk galaxies but fail at the ellipticals ones. It is remarkable that the RGGR model is doing fine in both cases.

\section{A brief review on RGGR}

The gravitational coupling parameter $G$ may  behave as a true constant in the far  IR limit, leading to standard General Relativity in such limit. Nevertheless, in the context of QFT in cuved space time, there is no proof on that. According to Refs. \cite{Shapiro:2004ch, Farina:2011me}, a certain logarithmic running of $G$  is a direct consequence of covariance and must hold in all loop orders.  Hence the situation is as follows: either there is no new gravitational effect induced by the renormalization group in the far infrared, or there are such deviations and the gravitational coupling runs as

\be
	 \beta_{G^{-1}} \equiv \mu \frac{dG^{-1}}{d \mu} = 2 \nu \,  \frac{M_{\mbox{\tiny Planck}}^{2}}{c \, \hbar} = 2 \nu G_0^{-1}.
	\label{betaG}
\ee
Equation (\ref{betaG}) leads to the logarithmically  varying $G(\mu)$ function,
\be
	\label{gmu}
	G(\mu) = \frac {G_0}{ 1 + \nu \ln(\mu^2/\mu_0^2)},
\ee
where $\mu_0$ is a reference scale introduced such that $G(\mu_0) =G_0 $. The constant $G_0$ is the gravitational constant as measured in the Solar System  (actually, there is no need to be very precise on where $G$ assumes the value of $G_0$, due to the smallness of the variation of $G$). The dimensionless constant $\nu$ is a phenomenological parameter which depends on the details of the quantum theory leading to eq. (\ref{gmu}). Since we have no means to compute the latter from first principles, its value should be fixed from observations. Even a small $\nu$ of about  $\sim 10^{-7}$ can lead to observational consequences at galactic scales. Note that the first possibility, namely of no new gravitational effects in the far infrared, corresponds to $\nu=0$.

The action for this model is simply the Einstein-Hilbert one in which $G$ appears inside the integral, namely,\footnote{We use the $(- + + +)$ space-time signature.}
\be
	S_{\mbox{\tiny RGGR}}[g] = \frac {c^3}{16 \pi }\int \frac {R  } G \, \sqrt{-g} \,  d^4x.
	\label{rggraction}
\ee
In the above, $G$ should be understood as an external scalar field that satisfies (\ref{gmu}).  Since for the problem of galaxy rotation curves  the cosmological constant effects are negligible, we have not written the $\Lambda$ term above. Nevertheless, for a complete cosmological picture, $\Lambda$ is necessary and it also runs covariantly with the RG flow of $G$ \cite{Shapiro:2004ch, Reuter:2007de}.

There is a simple procedure to map the solutions from the Einstein equations with the gravitational constant $G_0$  into RGGR solutions. In this review, we will proceed to find RGGR solutions via a conformal transformation of the Einstein-Hilbert action, and to this end first we write 
\be
	G = G_0 + \delta G,
\ee 
and we assume $\delta G / G_0 \ll 1$, which will be justified latter. Introducing the conformally related metric
\be
	\bar g_{\mu \nu} \equiv \frac {G_0}{G} g_{\mu \nu}, 
	\label{ct}
\ee
the RGGR action can be written as
\be
	S_{\mbox{\tiny RGGR}}[g] = S_{\mbox{\tiny EH}}[\bar g] + O(\delta G^2),
\ee
where $S_{\mbox{\tiny EH}}$ is the Einstein-Hilbert action with $G_0$ as the gravitational constant. The above suggest that the RGGR solutions can be generated from the Einstein equations solutions via the conformal transformation (\ref{ct}). Indeed, within a good approximation, one can check that this relation persists when comparing the RGGR equations of motion to the Einstein equations even in the presence of matter \cite{Rodrigues:2009vf}.

In the context of  galaxy kinematics, standard General Relativity gives essentially the same predictions of Newtonian gravity. In the weak field limit and for velocities much lower than that of light, the gravitational dynamics can be derived from the Newtonian potential, which is related to the metric by 
\be
	\bar g_{00} = - \left ( 1  + \frac {2 \Phi_\Newt}{c^2} \right ).
\ee
Hence, using eq. (\ref{ct}), the effective RGGR potential $\Phi$ is given by
\be
	\Phi = \Phi_\Newt + \frac {c^2}2 \frac{\delta G}{G_0}.
	\label{PhiRGGR}
\ee
An equivalent result can also be found from the geodesics  of a test particle\cite{Rodrigues:2009vf}. For weak gravitational fields $\Phi_\Newt/ c^2 \ll 1$ (with $\Phi_\Newt = 0$ at spatial infinity), hence even if  $\delta G/G_0 \ll 1$ eq. (\ref{PhiRGGR}) can lead to a significant departure from Newtonian gravity. 

In order to derive a test particle acceleration, we have to specify the proper energy scale $\mu$ for the problem setting in question, which is a time-independent gravitational phenomena in the weak field limit. This is a recent area of exploration of the renormalization group application, where the usual procedures for high energy scattering of particles cannot be applied straightforwardly. Previously to \cite{Rodrigues:2009vf} the selection of $\mu \propto 1/r$, where $r$ is the distance from a massive point, was repeatedly used, e.g. \cite{Reuter:2004nv,Dalvit:1994gf,Bertolami:1993mh,Goldman:1992qs, Shapiro:2004ch}. This identification adds a constant velocity proportional to $\nu$ to any rotation curve. Although it was pointed as an advantage due to the generation of ``flat rotation curves'' for galaxies, it introduced difficulties with the Tully-Fisher law \cite{Tully:1977fu}, the Newtonian limit, and the behavior of the galaxy rotation curve close to the galactic center, since there the behavior is closer to the expected one without dark matter. In \cite{Rodrigues:2009vf} we introduced a $\mu$ identification that seems better justified both from the theoretical and observational points of view. The characteristic weak-field gravitational energy scale does not comes from the geometric scaling $1/r$, but should be found from the Newtonian potential $\Phi_\Newt$, the latter is the field that characterizes gravity in such limit. Therefore,
\be
	\frac{\mu}{\mu_0} = f\( \frac{\Phi_N}{\Phi_0}\).
	\label{muf}
\ee
If $f$ would be a complicated function with dependence on diverse constants, that would lead to a theory with small (or null)  prediction power. The simplest assumption, $ \mu \propto \Phi_\Newt$,  leads to $\mu \propto 1/r$ in the large $r$ limit; which is unsatisfactory on observational grounds (bad Newtonian limit and correspondence to the Tully-Fisher law). One way to recover the Newtonian limit is to impose a suitable cut-off, but this rough procedure does not solves the Tully-Fisher issues \cite{Shapiro:2004ch}. Another one is to use \cite{Rodrigues:2009vf}
\be
	\frac {\mu}{\mu_0} =\left( \frac{\Phi_\Newt}{\Phi_0} \right)^\alpha,
	\label{muphi}
\ee
where $\Phi_0$ and $\alpha$ are constants. Apart from the condition $ \Phi_0 < 0$, in order to guarantee $\delta G/G_0 \ll 1$, the precise value of $\Phi_0$ is largely irrelevant for the dynamics, since $\Phi'(r)$ does not depends on $\Phi_0$. The relevant parameter is $\alpha$, which will be commented below.  The above energy scale setting (\ref{muphi}) was recently re-obtained from a more fundamental perspective \cite{Domazet:2010bk}, where a renormalization group scale-setting formalism is employed.

The parameter $\alpha$ is a phenomenological parameter that needs to depend on the mass of the system, and it must go to zero when the mass of the system goes to zero. This is necessary to have a good Newtonian limit. From the Tully-Fisher law, it is expected to increase monotonically with the increase of the mass of disk galaxies. In a recent paper, an upper bound on $\nu \alpha$ in the Solar System was derived \cite{Farina:2011me}. In galaxy systems, $\nu \alpha|_{\mbox{\tiny Galaxy}} \sim 10^{-7}$, while for the Solar System, whose mass is about $10^{-10}$ of that of a galaxy,  $\nu \alpha|_{\mbox{\tiny Solar System}} \lesssim 10^{-17}$. It shows that a linear increase on $ \alpha$ with the mass (ignoring possible dependences on the mass distribution) is sufficient to satisfy both the current upper bound from the Solar System and the results from galaxies. Actually, in Ref. \cite{Rodrigues:2012qm} it is shown that a close-to-linear dependence on the mass can also be found for elliptical galaxies by using the fundamental plane.

Once the $\mu$ identification is set, it is straightforward to find the rotation velocity for a static gravitational system sustained by its centripetal acceleration \cite{Rodrigues:2009vf},
\be
	V^2_{\mbox{\tiny RGGR}} \approx V^2_\Newt \left ( 1 - \frac {\nu \, \alpha  \, c^2} {\Phi_\Newt} \right ).
	\label{v2rggr}
\ee
Contrary to Newtonian gravity, the value of the Newtonian 
potential at a given point does play a significant role 
in this approach. This sounds odd from the perspective of 
Newtonian gravity, but this is not so 
from the General Relativity viewpoint, since the latter has no 
free zero point of energy. In particular, the Schwarzschild 
solution is not invariant under a constant shift of the 
potential.

Equation (\ref{v2rggr}) was essential for the derivation of  the RGGR galaxy rotation curves. Since elliptical galaxies are mainly supported by velocity dispersions (VD), the main equation for galaxy  kinematics in this case will not be eq.(\ref{v2rggr}), but the following expression for the projected (line-of-sight) VD (see also \cite{Mamon:2004xk}),
\be
	 \sigma_p^2(R) = \frac {2 G_0}{I(R)} \int_R^\infty K\(\frac r R\) \frac{\ell(r) M(r) } r dr,
	 \label{sigma_pK}
\ee
where
{\small
\ba
	 K(u) \equiv && \frac 12 u^{2 \beta - 1}\[ \(\frac 32 - \beta \) \sqrt \pi \frac{\Gamma(\beta - 1/2)}{\Gamma(\beta)} + \right. \\ 
	&& \left. + \beta B\(\frac 1{u^2}, \beta + \frac 12, \frac 12 \) -  B\(\frac 1{u^2}, \beta - \frac 12, \frac 12 \)\], \nonumber
\ea}
$B(x,a,b) = \int_0^x t^{a-1} (1-t^{b-1}) dt$ is the incomplete beta function, $\Gamma$ is the Gamma function, $r$ stands for the deprojected (spherical) radius, $R$ for the projected (line of sight) radius,  $I(R)$ is the luminosity intensity (its integral on an infinity surface gives the total luminosity $L$ of the galaxy),  $\ell(r)$ is the luminosity density (found from the deprojection of $I(R)$), $\beta$ is the anisotropy parameter (assumed constant inside each galaxy, it is zero if the galaxy has an isotropic VD profile) and $M(r)$ is the total (effective) mass of the system at the radius $r$. See Ref. \cite{Rodrigues:2012qm} and references therein for additional details. In the case of Newtonian gravity without dark matter $M(r)$ would stand, within a good approximation, as the internal stellar mass to radius $r$, i.e., $M_*(r)$.

For RGGR without dark matter, the total mass inside the radius $r$ is also the baryonic mass $M_*(r)$. Nevertheless, the non-Newtonian gravitational effects of RGGR for spherically symmetric systems can be understood from the Newtonian perspective as if the total mass was given by the following total effective mass \cite{Rodrigues:2012qm}
 \be
	M(r) = M_*(r) + M_{\mbox{\tiny RGGR}}(r),
\ee
with 
\be
	M_{\mbox{\tiny RGGR}} (r) = \frac{\alpha \nu c^2}{G_0} \frac{r}{1 + \frac{4 \pi r }{M_*(r)} \int_{r}^{\infty} \rho_*(r') r' dr'}.
	\label{MRGGR}
\ee

\section{NGC 4374 and DDO 47}

\subsection{NGC 4374}

NGC 4374 is a giant elliptical recently analyzed within RGGR \cite{Rodrigues:2012qm}.  Here we present a variation of the analysis presented in that reference, namely we here do not directly use the photometric data with the S\'ersic extension to model the stellar mass, instead we here only use the  S\'ersic model parameters that best fit the surface brightness of this galaxy. See Table \ref{N4374resultsTable} and Fig. \ref{N4374CosmoSul}. This is relevant to display that our results presented in \cite{Rodrigues:2012qm} are sufficiently robust to small changes on the baryonic mass content. 

\begin{table*}[htdp] 
\begin{center}
{\footnotesize 
\begin{tabular}{l c c c c c c}
\multicolumn{6}{c}{\emph{ \normalsize NGC 4374}}\\
\hline \hline
\multicolumn{6}{c}{\emph{RGGR without dark matter} } 				  \\ 
Stellar model $^{(1)}$		& $\alpha\,\nu \times 10^{7}$			& $\beta		$	 		&\ML 				&${\chi^{2}}$	&${\chir}$\\ \cline{1-6}
Full S\'ersic+$\beta_{[0]}$ 				&$15.2\pm2.4$				&0						&$3.94\pm0.56$		&20.1			&0.96\\
Full S\'ersic+$\beta_{[-1,1]}$				&$19.0^{+6.0}_{-6.3}$		&$0.57^{+0.35}_{-1.1}$		&$2.3^{+2.5}_{-2.3}$		&18.5			&0.92\\
Full S\'ersic+K.IMF+$\beta_{[0]}$			&$13.5\pm1.4$				&0						&$4.36\pm0.27$		&21.9			&0.99\\
Full S\'ersic+K.IMF+$\beta_{[-1,1]}$ 		&$13.7\pm{1.9}$			&$-0.20^{+0.41}_{-0.68}$		&$4.44\pm{0.39}$		&21.2			&1.0\\ 	

\hline \hline
\end{tabular}
\caption{\label{N4374resultsTable} \footnotesize NGC 4374 results for RGGR and assuming that the stellar mass profile is only given by a S\'ersic profile. This table extends a table on this galaxy in Ref.\cite{Rodrigues:2012qm}, see this reference for further details. (1) $\beta_{[0]}$ indicates isotropic VD, $\beta_{[-1,1]}$ indicates constant anisotropy with $\beta \in [-1,1]$, K.IMF is a reference to Kroupa IMF, and it means that the expected value of $\Upsilon_*$ was used (i.e., $4.5 \pm 1.0 M_\odot / L_{\odot, V}$) \cite{Rodrigues:2012qm}. Note that with this full S\'ersic profile, for this galaxy,  the RGGR fit is slightly better. The changes on the best fit parameters above, in regard to those found in Ref. \cite{Rodrigues:2012qm}, are well inside the 1$\sigma$ uncertainties.  Hence, small changes in the baryonic matter are indeed generating small changes on the model parameters. }}
\end{center}
\end{table*}

\begin{figure*}[thbp]
\begin{center}
	  \includegraphics[width=100mm]{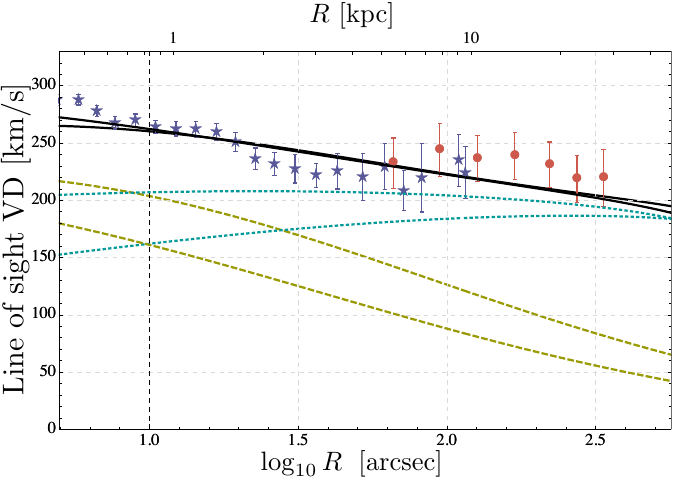}
\caption{Two NGC 4374 mass models with RGGR gravity and without dark matter. The curves refer to mass models  composed by the stellar component (given by a S\'ersic profile) and RGGR gravity. The black solid lines are the resulting VD for each model, the yellow dashed and blue dotted lines are respectively the stellar Newtonian and non-Newtonian contributions to the total VD. One of the models assumes isotropy ($\beta = 0$),  the second   assumes $\beta \in [-1,1]$. The vertical dashed line signs the radius above which the observational data is considered  for the fitting procedure (10 arcsec).}
\label{N4374CosmoSul}
\end{center}
\end{figure*}

In Ref. \cite{Rodrigues:2012qm} it is  shown hat MOND fits better the NGC 4374 observational data than Newtonian gravity without dark matter. However, it is still a poor fit, in particular since: $i$) There is a significant tendency towards a lower VD curve at large radii, tendency which is strongly enhanced once the fits consider the expected $\Upsilon_*$; $ii$) if the expected $\Upsilon_*$ is not used, the best fit is achieved for tangential anisotropy with  $\beta \le -1$.  Other  issues of MOND with the giant ellipticals can be found for instance in Ref.\cite{Gerhard:2000ck}. It was also shown that RGGR fit to the data is a satisfactory one and outperforms MOND in all the points above.

\subsection{DDO 47}

Here we present part of a new result from a working in progress\cite{FabrisFuture}. The dwarf disk galaxy DDO 47 is cited as a paradigmatic galaxy for testing dark matter effects \cite{2001AJ....121.3026W, 2003A&A...409...53S,2005ApJ...634L.145G}, and in particular its fits using the NFW profile vary from bad (using the two NFW halo parameters as free parameters) to a disaster (if one of the parameters is fixed in accordance with N-body simulations expectations) \cite{2005ApJ...634L.145G}.

Here we use the same data and conventions used in Ref.\cite{2005ApJ...634L.145G} for the baryonic matter and observed rotation curve, except for the gas surface density. For the latter, we directly use the gas surface density provided by Ref. \cite{2001AJ....121.3026W}, which leads to a gas rotation curve less smooth than the one presented in Ref.\cite{2005ApJ...634L.145G}. The results are not significantly sensitive to either of the gas profiles and are shown in Fig. \ref{DDO47CosmoSul}.

\begin{figure*}[thbp]
\begin{center}
	  \includegraphics[width=130mm]{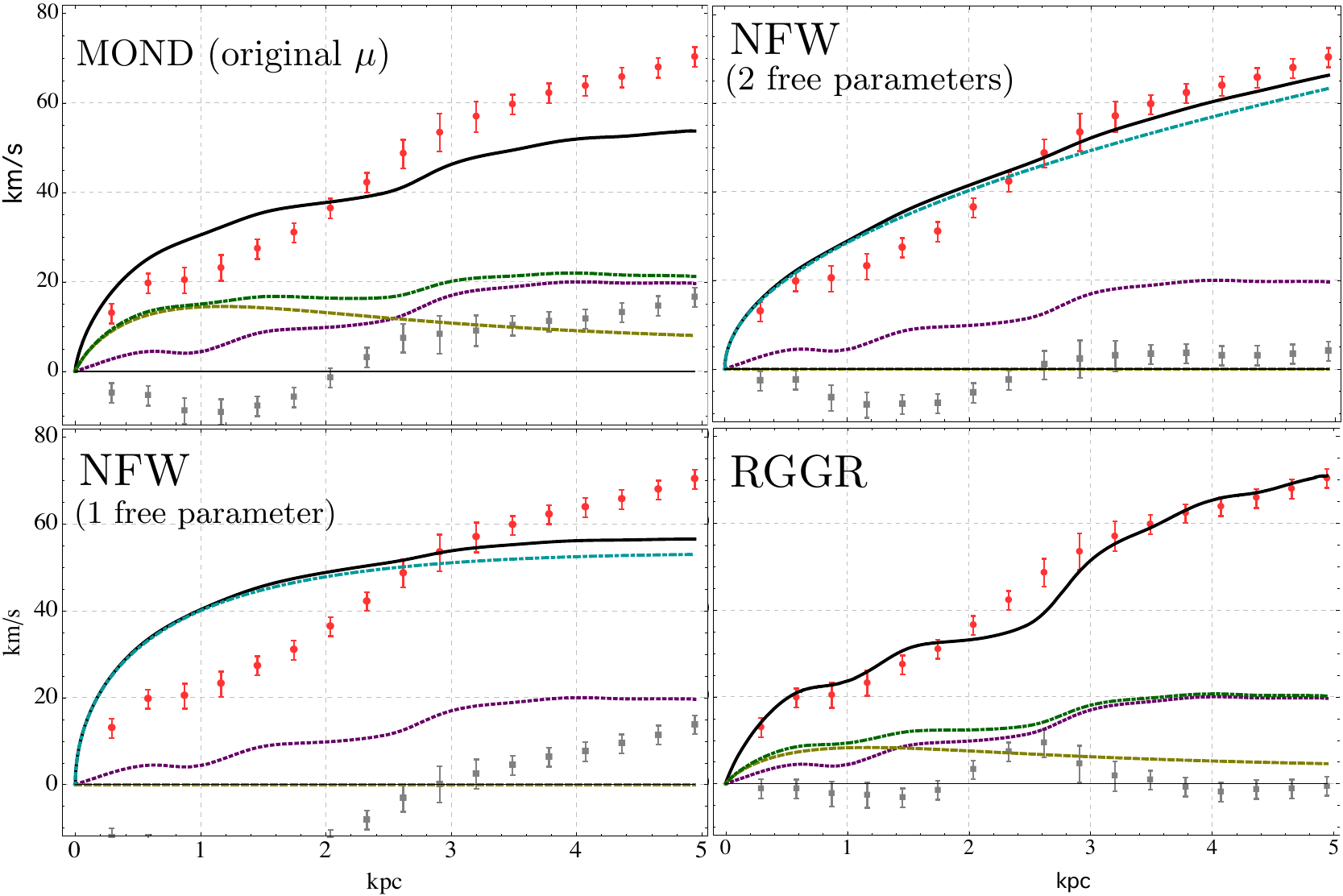}
\caption{DDO 47 rotation curve fits. The red dots and its error bars are the rotation curve observational data, the gray ones close to the abscissa are the residues of the fit. The solid black line for each model is its best fit rotation curve, the dashed yellow curves are the stellar rotation curves,  the dotted purple curve is the gas rotation curve, the dot-dashed green curve (shown in MOND and RGGR plots) is the resulting Newtonian, with no dark matter, rotation curve, and the blue dot-dashed curve present in the NFW plots is the dark matter halo contribution to the total rotation curve. Note that both of the NFW fits favor zero mass-to-light ratios, which is clearly wrong. We stress that the problems with the NFW halo fitting were not found here for the first time  (we only found them again from slightly different conventions), see Ref.\cite{2005ApJ...634L.145G} and references therein for further details. The result for RGGR above is a new one. For MOND, similar considerations appeared in the ArXiv version of the Ref.\cite{2005ApJ...634L.145G}.}
\label{DDO47CosmoSul}
\end{center}
\end{figure*}

\section{Conclusions}

In summary, RGGR is a model based on the theoretical possibility that the beta function of the gravitational coupling parameter $G$ may not be zero in the far infrared. Currently, there is no way to directly deduce this behavior from first principles, nevertheless eq. (\ref{betaG}) is a natural, if not unique, possibility that has appeared many times before in the context of Quantum Field Theory on curved space-time. This equation depends on a universal free parameter $\nu$ that can be constrained from experiments and observations, with $\nu = 0$ corresponding to standard General Relativity. The eq. (\ref{betaG}) also depends on an energy scale $\mu$, which should be related to the symmetries and physical interactions that are being evaluated. Considering gravitation effects in stationary systems with weak Newtonian potential ($\Phi_N /c^2\ll 1$) and slow particle velocities ($v/c \ll 1$), it is natural to use a relation of the type (\ref{muf}). In Ref. \cite{Rodrigues:2009vf} the relation (\ref{muphi}) was first proposed.

RGGR  without dark matter is a model with  one phenomenological free parameter ($\alpha$) which is capable of dealing with the kinematics of diverse galaxies. The $\alpha$  relation to other physical parameters we aim to understand soon \cite{FabrisFuture}.

\vspace{.2in}

\noindent
{\bf Acknowledgements }

DCR thanks the I CosmoSul organizers for the invitation. We thank Nicola Napolitano for a valuable remark on the NGC 4374 stellar mass model that we use here, and Paolo Salucci for kindly providing the data of DDO 47. DCR and JCF thank CNPq and FAPES for partial financial support. PLO thanks CAPES for financial support. The work of I.Sh. has been partially supported by CNPq, FAPEMIG and ICTP.

\bibliographystyle{aipnum4-1}

\bibliography{/Users/Davi/Desktop/Works/bibdavi2010}{}

\end{document}